\documentclass[fdp,a4paper,fleqn%
]{w-art}
\usepackage{times,cite,w-thm}
\theoremstyle{plain}

\theoremstyle{definition}

\usepackage[]{graphicx}
\begin{document}
\par\hfill DFTT 1/2011

\par\hfill ITP-UU-11/02

\DOIsuffix{theDOIsuffix}
\Volume{55}
\Month{01}
\Year{2007}
\pagespan{1}{}



\title[Hydrodynamics of holographic flavored plasmas]{Hydrodynamics of holographic flavored plasmas}


\author[F. Bigazzi]{Francesco Bigazzi\inst{1,}\footnote{bigazzi@fi.infn.it}}
\address[\inst{1}]{Dipartimento di Fisica e Astronomia, Universit\'a di Firenze and INFN Sezione di Firenze; Via G. Sansone 1, I-50019 Sesto Fiorentino (Firenze), Italy.}
\author[A. L. Cotrone]{Aldo L. Cotrone\inst{2,}%
  \footnote{cotrone@to.infn.it \quad - \quad  Speaker}}
\address[\inst{2}]{Dipartimento di Fisica Teorica, Universit\`a di Torino and INFN Sezione di Torino;
Via P. Giuria 1, I-10125 Torino, Italy.}
\author[J. Tarr\'\i o]{Javier Tarr\'\i o\inst{3,}\footnote{l.j.tarriobarreiro@uu.nl}}
\address[\inst{3}]{Institute for Theoretical Physics, Universiteit Utrecht, 3584 CE, Utrecht, The Netherlands.}
   \dedicatory{F. B. and A. L. C. would like to thank the Italian students, parents, teachers and scientists for their activity in support of public education and research.}
\begin{abstract}
We provide a short review of the hydrodynamical properties of a class of strongly
coupled thermal gauge theories, coupled to massless fundamental flavor fields, having a dual supergravity description. The analysis is valid for more general classes of nearly conformal holographic plasmas, where the breaking of conformality is driven by marginally (ir)relevant deformations. 
\end{abstract}
\maketitle                   





\section{Introduction}

The analysis of the experimental data associated to the production of the quark-gluon plasma (QGP) in heavy ion collisions at RHIC (Brookhaven, USA) and LHC (CERN) is interpreted, despite the large errors involved, as an evidence that this state of matter is a strongly interacting fluid at high temperature ($\sim200$ MeV), composed of deconfined adjoint (gluons) and fundamental (quarks) matter. The QGP is supposed to have existed in the immediate moments after the big bang, hence the importance to understand its behaviour.

Unfortunately, due to the strong nature of the interaction, the well-known perturbative methods of QCD are not sufficient to study the QGP. Lattice calculations proved a valuable tool, however they are not well suited to study real-time properties of the system. These properties include the transport coefficients which govern the hydrodynamic behaviour at long distances and times as compared to the inverse temperature. Were these coefficients known, especially the viscosities and relaxation times, we would be able to run computer simulations and compare the theoretical predictions with the observed experimental behaviour.

The AdS/CFT correspondence \cite{malda, gkp, witten} exploits the holographic principle to study strongly coupled $D$-dimensional conformal quantum field theories by means of dual $D+1$-dimensional gravitational models in asymptotically anti-de Sitter spaces. Within this correspondence (which can be extended to many non-conformal setups as well) a thermal gauge theory is associated with a black hole background. Each fluid mode in the plasma has a corresponding gravity mode, whose fluctuations, governed by gravity equations, can be used to obtain retarded correlators, from where we can obtain the transport coefficients. The main ingredient in the AdS/CFT correspondence is the relation
\begin{equation}
\langle e^{-\int \phi_0 {\cal O}} \rangle = e^{- S_{gravity}(\phi_0)} \, ,
\end{equation}
where the source $\phi_0$ of the field theory operator $\cal O$ is identified with the value of the dual gravitational mode at the AdS boundary $\phi_0 = \lim_{r\to\infty}\phi(r)$, and $r$ is the AdS radial coordinate. The correspondence is actually a limit of a conjectured more general equivalence between quantum field theories and higher dimensional string models having a gravitational description at low energy and weak coupling.

\section{Hydrodynamics from AdS/CFT}

At long distances and times as compared to the inverse temperature, field theories admit a hydrodynamic description dictated by the conservation of energy and momentum. This hydrodynamic description can be organized in a derivative expansion. Up to second derivatives the expansion of the energy-momentum tensor for a relativistic uncharged fluid reads\cite{baier,romatschke}
\begin{equation}\label{enmom}
T^{\mu\nu}=\varepsilon u^\mu u^\nu + p  \Delta^{\mu\nu} + \pi^{\mu\nu} + \Delta^{\mu\nu}\Pi   \, ,
\end{equation}
where $\varepsilon$ is the energy density, $u^\mu$ the velocity field, $p(\varepsilon)$ the pressure, $\Delta^{\mu\nu} = h^{\mu\nu}+u^\mu u^\nu$ with $h^{\mu\nu}$ the $4$-dimensional metric and
\begin{eqnarray}
\pi^{\mu\nu} &=&- \eta \sigma^{\mu\nu} +\eta \tau_\pi \Bigl[\langle D \sigma^{\mu\nu}\rangle + \frac{\nabla \cdot u}{3}\sigma^{\mu\nu} \Bigr] + \kappa \Bigl[ R^{<\mu\nu>}-2 u_\alpha u_\beta R^{\alpha <\mu\nu> \beta} \Bigr] \nonumber \\
&& + \lambda_1 \sigma^{<\mu}_{\lambda} \sigma^{\nu>\lambda} + \lambda_2  \sigma^{<\mu}_{\lambda} \Omega^{\nu>\lambda} + \lambda_3 \Omega^{<\mu}_{\lambda} \Omega^{\nu>\lambda}  + \kappa^* 2u_\alpha u_\beta R^{\alpha <\mu\nu> \beta}  \nonumber \\
&& + \eta \tau_\pi^* \frac{\nabla \cdot u}{3}\sigma^{\mu\nu} + \lambda_4 \nabla^{<\mu} \log{s}  \nabla^{\nu >} \log{s} \, ,
\end{eqnarray}
\begin{eqnarray}
\Pi &=&- \zeta (\nabla \cdot u) + \zeta \tau_\Pi  D  (\nabla \cdot u) + \xi_1 \sigma^{\mu\nu}\sigma_{\mu\nu}+ \xi_2 (\nabla \cdot u)^2 + \xi_3 \Omega^{\mu\nu}\Omega_{\mu\nu} \nonumber \\
&&  + \xi_4 \nabla_{\mu}^{\perp} \log{s} \nabla^{\mu}_{\perp} \log{s}+ \xi_5 R + \xi_6  u^\alpha u^\beta R_{\alpha \beta}\, . 
\end{eqnarray}
We refer the reader to \cite{romatschke} for the precise definitions of the structures in these formulas, which will not be necessary for the rest of this note. The shear viscosity $\eta$ and the second order coefficients $\tau_\pi$ (``shear" relaxation time), $\kappa$, $\lambda_1, \lambda_2, \lambda_3$ are the only ones defined in conformal fluids. All other coefficients, i.e. the bulk viscosity $\zeta$ and the second order coefficients $\kappa^*, \tau_\pi^*, \lambda_4, \tau_\Pi$ (``bulk" relaxation time), $\xi_1,   \xi_2,  \xi_3,  \xi_4,  \xi_5,  \xi_6$, are only defined in non-conformal plasmas.

Holographic methods allow to extract these transport coefficients in classes of strongly coupled plasmas having a dual gravity description. Moreover, in the regime where higher derivative corrections to the gravity action can be neglected, the corresponding plasmas display some relevant universal features. For example, they all have the same shear viscosity over entropy density ratio, $\eta/s=1/4\pi$, as the ${\cal N}=4$ supersymmetric Yang-Mills (SYM) plasma \cite{pss}. Remarkably this ratio is compatible with the one which can be deduced for the QGP at RHIC and LHC. This raises the hope that, at least in some limits, holographic results (despite strictly valid for theories still quite far from QCD), can be used as benchmarks for realistic simulations of real-time properties of the QGP. 

A sketch of the relevant holographic methods is as follows. Consider a fluid moving along one (say, $z$) of the 3 spatial directions $x,y,z$. For any field $\psi$ on the dual gravity background, consider fluctuations of the form $\exp(-i \omega t+ i q z)\psi(r)$, with $\omega$ and $q$ frequency and momentum. The fluctuations $\psi(r)$ are classified according to their transformation under $SO(2)_{x-y}$. Solving the equations of motions (with suitably chosen boundary conditions) for the fluctuations, one can get the dispersion relations and thus deduce the transport coefficients, taking into account general expressions like
\begin{equation}\label{vecdiff2}
\omega = c_s q - i \Gamma q^2 + \frac{\Gamma}{c_s}\Bigl(c_s^2\tau^{eff}-\frac{\Gamma}{2}\Bigr)q^3 + {\cal O}(q^4)\qquad {\rm where} \qquad \Gamma=\frac{\eta}{sT} \left( \frac{2}{3} + \frac{\zeta}{2\eta} \right)\,, 
\end{equation}
which holds for the scalar hydrodynamic modes \cite{baier,romatschke}.\footnote{Here $T$ and $c_s$ are the temperature and speed of sound of the plasma. Finally, $\tau^{eff}$ is an ``effective relaxation time''.}
Another source of information comes from the study of retarded correlators of the stress-energy tensor. For the tensorial mode \cite{baier,romatschke}
\begin{equation}\label{retcorr}
G_R^{xy,xy}=p-i \eta \omega + \Bigl( \eta \tau_\pi -\frac{\kappa}{2} +\kappa^* \Bigr)\omega^2 -\frac{\kappa}{2}q^2 + {\cal O}(q^3,\omega^3)\, ,
\end{equation}
where $p$ is the pressure. The holographic computation of these correlators gives direct access to the related transport coefficients.

\section{A flavored ${\cal N}=4$ SYM plasma}
Theories like (thermal) ${\cal N}=4$ $SU(N_c)$ SYM, which has a dual $AdS_5\times S^5$ (black hole) description, do not have matter fields transforming in the fundamental representation. The inclusion of fundamental matter has a precise counterpart in the dual string/gravity setup. It amounts on adding extended sources (like $N_f$ ``flavor" D7-branes) on the background. In the 't Hooft limit (where $N_c\to\infty$ whereas the 't Hooft coupling $\lambda$ and $N_f$ are kept fixed), the branes can be treated as probes \cite{kk} and thus do not deform the original background. This corresponds to taking the quenched approximation for the flavor fields in the dual gauge theory. Going beyond this approximation requires accounting for the backreaction of the flavor branes on the background. This is a difficult task in general, since the branes (which have codimension 2 in the D7 case and thus are localized at some angles of the $5$-sphere $S^5$) enter as delta function sources in the supegravity equations of motion and Bianchi identities. This gives rise to a set of partial differential equations to be solved for.

In \cite{noncritical} a method named ``smearing technique'' was introduced. This method is appropriate in the Veneziano limit in which $N_c\to\infty$ and $N_f\to \infty$ with their ratio fixed. Instead of considering $N_f$ localized branes one homogeneously distributes them in the transverse space, in such a way to replace delta function sources with a density distribution 2-form and to recover (most of) the isometries of the original unflavored background. In this way one often has to solve just ordinary differential equations in a radial variable (see \cite{npr} for a review).

In \cite{Bigazzi:2009bk} this method was applied to thermal ${\cal N}=4$ SYM with massless fundamental hypermultiplets (and then extended to more general flavored quivers), finding a solution which takes into account the D7-brane backreaction in a perturbative expansion in the parameter
$
\epsilon_h \propto \lambda_h \frac{N_f}{N_c} \, ,
$
with $\lambda_h$ the 't Hooft coupling at the energy scale set by the temperature $T$. This parameter would weigh the internal quark loops in a perturbative expansion of, say, gluon polarization diagrams. In the string setup, it has to be taken very small in order for the gravity description to be reliable. The solution in \cite{Bigazzi:2009bk} was analytically found to order $\epsilon_h^2$. It is relevant to notice that the class of flavored plasmas here considered are examples of non-conformal models. The breaking of conformal invariance, driven by quantum effects since the flavors are massless, is precisely encoded by the running of the beta function for $\epsilon_h$:
$
\ T \frac{d \epsilon_h}{d T} = \epsilon_h^2+ {\cal O}(\epsilon_h^3).
$

The gravity solution in \cite{Bigazzi:2009bk} has a warped black hole metric of the form 
\begin{equation}
ds^2 = \frac{r^2}{R^2} \left[-\left(1-\frac{r_h^4}{r^4}\right)\,dt^2 + dx_idx_i\right] + \frac{R^2}{r^2}\left[\frac{S^8F^2}{1-\frac{r_h^4}{r^4}}dr^2 + r^2 \left(S^2 ds_{KE}^2 + F^2 (d\tau + A_{KE})^2\right)\right]\,,
\label{thebac}
\end{equation}
where $r_h$ is the horizon radius. The flavor brane backreaction is accounted for by the functions $S(r),F(r)$ and the metric of the original $S^5$ is expressed as a $U(1)$ fibration over a K\"ahler-Einstein base $CP^2$ ($dA_{KE}/2=J_{KE}$ is the K\"ahler form of the four-dimensional base of $S^5$). Moreover
\begin{eqnarray}
F &=& 1 - \frac{\epsilon_h}{24} + \frac{17}{1152}\epsilon_h^2 -\frac{\epsilon_h^2}{24}\log\frac{r}{r_h} \,,\\
S &=& 1 + \frac{\epsilon_h}{24} + \frac{1}{128}\epsilon_h^2 + \frac{\epsilon_h^2}{24}\log\frac{r}{r_h}\,,\\
\Phi &=& \Phi_h + \epsilon_h \log\frac{r}{r_h} + \frac{\epsilon_h^2}{6} \log\frac{r}{r_h} + \frac{\epsilon_h^2}{2} \log^2\frac{r}{r_h} + \frac{\epsilon_h^2}{16} Li_2\left(1-\frac{r_h^4}{r^4}\right)\,,
\label{simple}
\end{eqnarray}
where we have also included the running dilaton $\Phi$. The solution contains $F_5$ and $F_1$ Ramond-Ramond field strengths too. We refer to \cite{Bigazzi:2009bk} for details and comments on the UV behaviour of the solution.

The solution described above allows us to study a number of effects of dynamical flavors in a strongly coupled thermal theory in a completely controllable setting.
Some thermodynamic quantities (entropy density $s$, energy density $\varepsilon$,  free energy density $f$, speed of sound $c_s$)  are \cite{Bigazzi:2009bk}
\begin{eqnarray}
s&=&\frac12 \pi^2 N_c^2 T^3 \left[1+\frac12 \epsilon_h 
+\frac{7}{24}\epsilon_h^2 \right]\, ,\\
f &=&-p=-\frac18 \pi^2
 N_c^2 T^4 \left[1+\frac12 \epsilon_h +\frac16 \epsilon_h^2\right]\, , \\
\varepsilon-3p&=&\frac{1}{16}\pi^2 N_c^2 T^4 \epsilon_h^2\, ,\\
c_s^2 &=&  \frac13 \left[1-\frac{1}{6} \epsilon_h^2\right]\, .
\end{eqnarray}
The transport coefficients up to ${\cal O}(\epsilon_h^2)$ obtained by studying the gravitational fluctuations, as sketched in the previous section, are \cite{noi} 
\begin{eqnarray}\label{resultbulk}
\frac{\zeta}{\eta}&=&\frac19 \epsilon^2_h\,,\\ \label{resulttau}
\tau^{eff}T&=&\tau_{\pi,0}T_{0} + \frac{16-\pi^2}{128\pi}\epsilon_h^2\,,\\ \label{resultk}
\frac{T^2}{p}\kappa &=&\frac{T_{0}^2}{p_{0}}\kappa_{0}\,,\\\label{resultkstar}
\frac{T^2}{p}(\kappa^*+\eta\tau_\pi)&=&\frac{T_{0}^2}{p_{0}}\eta_{0}\tau_{\pi,0} + \frac{T_{0}^2}{p_{0}}\eta_{0}\Bigl(\frac{\tau_{\pi,0}}{8}-\frac{1}{8\pi T_{0}}\Bigr)\epsilon_h^2\,,
\end{eqnarray}
where $ \tau_{\pi,0}T_{0}=\frac{2-\log{2}}{2\pi} $ and $ \frac{T_{0}^2}{p_{0}}\kappa_{0}=\frac{1}{\pi^2}\, $, are the corresponding values in the conformal plasmas.

\section{Hydrodynamics from AdS/CFT revisited}
There is a simple way to obtain, holographically, all the second order transport coefficients in the above flavored plasmas, avoiding the explicit study of fluctuating modes and correlators. The flavored ${\cal N}=4$ plasma has a dual effective $5$-dimensional description in terms of a metric and three scalars \cite{Benini:2006hh,noi}. One of these scalars is the dilaton. The others describe the volume of the compact deformed $S^5$ and the squashing between the fiber and the base. The corresponding field theory operators have dimensions $\Delta=4,8,6$ at the unflavored conformal fixed point. Thus, giving a non trivial profile to the dilaton around the AdS background corresponds to turning on a (marginally irrelevant) deformation in the field theory. The other scalars, instead, would drive irrelevant deformations.

At order $\epsilon_h^2$ the breaking of conformality can be accounted for just by the dilaton, so that the $5D$ model reduces effectively to a single scalar one. Crucially, the latter is in the Chamblin-Reall class \cite{Chamblin:1999ya}, already studied in \cite{Gubser:2008sz} as for some hydrodynamical properties. In more generality, Chamblin-Reall models (which are characterized by a simple exponential potential for the scalar field) provide good effective holographic descriptions, at leading order in the deformation, for classes of strongly coupled conformal gauge theories slightly deformed by marginally (ir)relevant operators \cite{stima}. 

Quite crucially, for the Chamblin-Reall theories, all the hydrodynamic transport coefficients up to second order can be extracted \cite{stima} using the results in \cite{Kanitscheider:2009as}.  
With the definition
\begin{equation}\label{Delta}
\delta \equiv 1-3c_s^2\,, 
\end{equation}
where $c_s$ is the speed of sound, and referring to the hydrodynamic stress-energy tensor in (\ref{enmom}),
the transport coefficients are given in Table \ref{relations}.\footnote{The flavored ${\cal N}=4$ SYM plasma has these same coefficients with $\delta=\epsilon_h^2/6$.}
\begin{table}[h]\label{relations}
\begin{center}
\caption{The transport coefficients at leading order in the conformality deformation parameter $\delta \equiv 1-3c_s^2$.}\label{relations}
\begin{tabular}{||c|c||c|c||c|c||} 
\hline
 & & & & & \\
$ \frac{\eta}{s} $ & $\frac{1}{4\pi}$ &  $T\tau_{\pi}  $  & $ \frac{2-\log{2}}{2\pi} + \frac{3(16-\pi^2)}{64\pi}\delta $  & $ \frac{T\kappa}{s} $  &  $  \frac{1}{4\pi^2}\Bigl(1-\frac34 \delta \Bigr) $  \\
 & & & & & \\
\hline \hline
 & & & & & \\  
$\frac{T \lambda_1}{s}  $ & $\frac{1}{8\pi^2}\Bigl(1+\frac34 \delta \Bigr) $ & $\frac{T \lambda_2}{s} $ & $-\frac{1}{4\pi^2}\Bigl( \log{2}+\frac{3\pi^2}{32}\delta \Bigr) $ & $\frac{T \lambda_3}{s} $ & $0 $ \\ 
 & & & & & \\
\hline \hline
 & & & & & \\
$\frac{T\kappa^*}{s} $ & $-\frac{3}{8\pi^2}\delta $ & $T\tau_{\pi}^* $ & $-\frac{2-\log{2}}{2\pi}\delta $ & $\frac{T \lambda_4}{s}  $ & $0 $ \\
 & & & & & \\
\hline \hline
 & & & & & \\
$\frac{\zeta}{\eta} $ & $\frac23 \delta $ & $T\tau_{\Pi} $ & $\frac{2-\log{2}}{2\pi} $ & $\frac{T \xi_{1}}{s} $ & $\frac{1}{24\pi^2}\delta $ \\
 & & & & & \\
\hline \hline
 & & & & & \\
$ \frac{T \xi_{2}}{s} $ & $\frac{2-\log{2}}{36\pi^2}\delta $ & $\frac{T \xi_{3}}{s} $ & $0 $ & $\frac{T \xi_{4}}{s} $ & $0 $ \\
 & & & & & \\
\hline \hline
 & & & & & \\
$\frac{T \xi_{5}}{s} $ & $\frac{1}{12\pi^2}\delta $ & $\frac{T \xi_{6}}{s} $ & $\frac{1}{4\pi^2}\delta $ & & \\
 & & & & & \\
\hline
\end{tabular} 
\end{center}
\end{table} 
Considering the difficulty of dealing with such coefficients in QCD, this information\footnote{In particular, the behavior with the temperature and the speed of sound of the shear and bulk relaxation times $\tau_\pi, \tau_\Pi$ is both potentially relevant and unexpected.} could be useful in numerical simulations of the hydrodynamic evolution of the QGP, provided at some stage of its thermalization (well above the critical temperature $T_c$ for deconfinement) it can be approximated by a small deformation of a conformal plasma in the class describe above. Actually, some of the thermodynamical properties of the QGP, as deduced from the lattice, in the temperature window $1.5 T_c \leq T \leq 4 T_c$ (relevant at RHIC and LHC), suggest that the QGP can be treated as a nearly conformal system. In order to provide a numerical example, taking $c_s^2\sim 0.26$ at $T\sim 1.5\,T_c$ as sensible estimate from lattice studies for the current RHIC experiment \cite{Katz:2005br1,Katz:2005br2,nuovo}, we would get the results collected in Table 2 (updating the ones in \cite{stima}).
\begin{table}[h]
\begin{center}
\caption{The transport coefficients at $T\sim 1.5\,T_c$ and $c_s^2 \sim 0.26$.}
\begin{tabular}{||c|c||c|c||c|c||}
\hline
 & & & & & \\
$ \frac{\eta}{s} $ & $\frac{1}{4\pi}$ &  $T\tau_{\pi}  $  & $0.228 $  & $ \frac{T\kappa}{s} $  &  $0.021 $  \\
 & & & & & \\
\hline \hline
 & & & & & \\  
$\frac{T \lambda_1}{s}  $ & $0.015 $ & $\frac{T \lambda_2}{s} $ & $-0.023 $ & $\frac{T \lambda_3}{s} $ & $0 $ \\ 
 & & & & & \\
\hline \hline
 & & & & & \\
$\frac{T\kappa^*}{s} $ & $-0.008 $ & $T\tau_{\pi}^* $ & $-0.046 $ & $\frac{T \lambda_4}{s}  $ & $0 $ \\
 & & & & & \\
\hline \hline
 & & & & & \\
$\frac{\zeta}{\eta} $ & $0.147 $ & $T\tau_{\Pi} $ & $0.208 $ & $\frac{T \xi_{1}}{s} $ & $0.001 $ \\
 & & & & & \\
\hline \hline
 & & & & & \\
$ \frac{T \xi_{2}}{s} $ & $0.001 $ & $\frac{T \xi_{3}}{s} $ & $0 $ & $\frac{T \xi_{4}}{s} $ & $0 $ \\
 & & & & & \\
\hline \hline
 & & & & & \\
$\frac{T \xi_{5}}{s} $ & $0.002 $ & $\frac{T \xi_{6}}{s} $ & $0.006 $ & & \\
 & & & & & \\
\hline
\end{tabular} 
\end{center}
\end{table}
\begin{acknowledgement}We thank A. Paredes and C. Ratti for useful observations.
F.B. and A.L.C. have received funding from the European
Community Seventh Framework Programme (FP7/2007-2013) under grant agreement
n. 253937 and 253534 respectively. J.T. has been supported by the Netherlands Organization for Scientific Research (NWO) under the FOM Foundation research program. J.T. is thankful to the Front of Galician-speaking Scientists for encouragement.
\end{acknowledgement}


\begin{thebibliography}{[1]}

\bibitem{malda}
  J.~M.~Maldacena,
  Adv.\ Theor.\ Math.\ Phys.\  {\bf 2}, 231 (1998)
  [Int.\ J.\ Theor.\ Phys.\  {\bf 38}, 1113 (1999)]
  [arXiv:hep-th/9711200].

\bibitem{gkp}
  S.~S.~Gubser, I.~R.~Klebanov and A.~M.~Polyakov,
  Phys.\ Lett.\  B {\bf 428}, 105 (1998)
  [arXiv:hep-th/9802109].

\bibitem{witten}
  E.~Witten,
  Adv.\ Theor.\ Math.\ Phys.\  {\bf 2}, 253 (1998)
  [arXiv:hep-th/9802150].

\bibitem{baier}
  R.~Baier, P.~Romatschke, D.~T.~Son, A.~O.~Starinets and M.~A.~Stephanov,
  JHEP {\bf 0804}, 100 (2008)
  [arXiv:0712.2451 [hep-th]].
S.~Bhattacharyya, V.~E.~Hubeny, S.~Minwalla and M.~Rangamani,
  JHEP {\bf 0802}, 045 (2008)
  [arXiv:0712.2456 [hep-th]].
 M.~Natsuume and T.~Okamura,
  Phys.\ Rev.\  D {\bf 77}, 066014 (2008)
  [Erratum-ibid.\  D {\bf 78}, 089902 (2008)]
  [arXiv:0712.2916 [hep-th]].

\bibitem{romatschke}
  P.~Romatschke,
Class.\ Quant.\ Grav.\  {\bf 27}, 025006 (2010)
  [arXiv:0906.4787 [hep-th]].
    
\bibitem{pss}G.~Policastro, D.~T.~Son and A.~O.~Starinets,
  Phys.\ Rev.\ Lett.\  {\bf 87}, 081601 (2001)
  [arXiv:hep-th/0104066].

\bibitem{kk}
  A.~Karch and E.~Katz,
  JHEP {\bf 0206}, 043 (2002)
  [arXiv:hep-th/0205236].

\bibitem{noncritical}
  F.~Bigazzi, R.~Casero, A.~L.~Cotrone, E.~Kiritsis and A.~Paredes,
  JHEP {\bf 0510}, 012 (2005)
  [hep-th/0505140].

\bibitem{npr}
  C.~Nunez, A.~Paredes and A.~V.~Ramallo,
  Adv.\ High Energy Phys.\  {\bf 2010}, 196714 (2010)
  [arXiv:1002.1088 [hep-th]].

\bibitem{Bigazzi:2009bk}
  F.~Bigazzi, A.~L.~Cotrone, J.~Mas, A.~Paredes, A.~V.~Ramallo and J.~Tarrio,
  JHEP {\bf 0911}, 117 (2009)
  [arXiv:0909.2865 [hep-th]].

\bibitem{noi}F.~Bigazzi, A.~L.~Cotrone and J.~Tarrio,
 JHEP {\bf 1002}, 083 (2010)
  [arXiv:0912.3256 [hep-th]].

\bibitem{Benini:2006hh}
 F.~Benini, F.~Canoura, S.~Cremonesi, C.~Nunez and A.~V.~Ramallo,
 JHEP {\bf 0702}, 090 (2007)
 [arXiv:hep-th/0612118].

\bibitem{Chamblin:1999ya}
  H.~A.~Chamblin and H.~S.~Reall,
  Nucl.\ Phys.\  {\bf B562}, 133-157 (1999)
  [hep-th/9903225].

\bibitem{Gubser:2008sz}
  S.~S.~Gubser, S.~S.~Pufu and F.~D.~Rocha,
  JHEP {\bf 0808}, 085 (2008)
  [arXiv:0806.0407 [hep-th]].

\bibitem{stima}F.~Bigazzi and A.~L.~Cotrone,
  JHEP {\bf 1008}, 128 (2010)
  [arXiv:1006.4634 [hep-ph]].

\bibitem{Kanitscheider:2009as}
  I.~Kanitscheider and K.~Skenderis,
  JHEP {\bf 0904}, 062 (2009)
  [arXiv:0901.1487 [hep-th]].
  
  \bibitem{Katz:2005br1}
Y.~Aoki, Z.~Fodor, S.~D.~Katz and K.~K.~Szabo,
  JHEP {\bf 0601}, 089 (2006)
  [arXiv:hep-lat/0510084].
  
  \bibitem{Katz:2005br2}
  S.~D.~Katz,
  Nucl.\ Phys.\  A {\bf 774}, 159 (2006)
  [arXiv:hep-ph/0511166].

\bibitem{nuovo}S.~Borsanyi {\it et al.},
  JHEP {\bf 1011}, 077 (2010)
  [arXiv:1007.2580 [hep-lat]].


\end{thebibliography}
\end{document}